%% file: arxiv.tex
\documentclass[superscriptaddress,aps,prmaterials,twocolumn,showpacs,showkeys,notitlepage,amsmath, amssymb,floatfix,citeautoscript]{revtex4-2}

\usepackage{graphicx}
\usepackage{nicefrac}
\usepackage{amsfonts}
\usepackage[flushleft]{threeparttable}

\usepackage{subfigure}
\usepackage{multirow} 
\usepackage{tabularx} 
\usepackage{array}
\usepackage[resetlabels,labeled]{multibib}
\newcites{SM}{Supplemental Material}

\usepackage{xcolor}
\usepackage{soul}
\usepackage{bm}

\usepackage{url}

\usepackage{booktabs}
\usepackage{threeparttable}

\newcommand{\FP}{Fe$_2$P}
\newcommand{\msr}{$\mu$SR}

\newcommand{\WK}{WIEN2k}
\newcommand{\elk}{Elk}

\newcommand{\commentout}[1]{}

\newcommand{\editor}[2]{%
  \expandafter\newcommand\csname #1note\endcsname[1]{%
    \textcolor{#2}{(\textbf{#1:} ##1)}}%
  \expandafter\newcommand\csname #1\endcsname[1]{%
    \textcolor{#2}{##1}}%
  \expandafter\newcommand\csname #1cancel\endcsname[1]{%
    \textcolor{#2}{\sout{##1}}}%
  \expandafter\newcommand\csname #1change\endcsname[2]{%
    \textcolor{#2}{\sout{##1} ##2}}%
  \newenvironment{#1text}{\color{#2}}{\color{black}}
}

\editor{PB}{red}
\newcommand{\papertitle}{\emph{Ab initio} modeling and experimental investigation of \FP{} by DFT and spin spectroscopies}

\begin{document}
\title{\papertitle}

\author{Pietro Bonf\`{a}}
\email[]{pietro.bonfa@unipr.it}
\affiliation{Department of Mathematical, Physical and Computer Sciences, University of Parma, 43124 Parma, Italy}
\affiliation{Centro S3, {CNR-Istituto} Nanoscienze, 41125 Modena, Italy}

\author{Muhammad Maikudi Isah}
\affiliation{Department of Mathematical, Physical and Computer Sciences, University of Parma, 43124 Parma, Italy}

\author{Benjamin A. Frandsen}
\affiliation{ %
	Department of Physics and Astronomy, Brigham Young University, Provo, Utah 84602, USA.
} %

\author{Ethan J. Gibson}
\affiliation{ %
    Department of Physics and Astronomy, Brigham Young University, Provo, Utah 84602, USA.
} %

\author{Ekkes Br\"uck}
\affiliation{Fundamental Aspects of Materials and Energy (FAME), Faculty of Applied Sciences, Delft University of Technology,
  2629 JB Delft, The Netherlands}

\author{Ifeanyi John Onuorah}
\affiliation{Department of Mathematical, Physical and Computer Sciences, University of Parma, 43124 Parma, Italy}

\author{Roberto De Renzi}
\affiliation{Department of Mathematical, Physical and Computer Sciences, University of Parma, 43124 Parma, Italy}

\author{Giuseppe Allodi}
\affiliation{Department of Mathematical, Physical and Computer Sciences, University of Parma, 43124 Parma, Italy}

\begin{abstract}
\FP{} alloys have been identified as promising candidates for magnetic refrigeration at room-temperature and for custom magnetostatic applications.
The intent of this study is to accurately characterize the magnetic ground state of the parent compound, \FP, with two spectroscopic techniques, $\mu$SR and NMR, in order to provide solid bases for further experimental analysis of \FP-type transition metal based alloys. We perform zero applied field measurements using both techniques below the ferromagnetic transition $T_C=220$~K. The experimental results are reproduced and interpreted using first principles simulations validating this approach for quantitative estimates in alloys of interest for technological applications.

\end{abstract}

\maketitle

\paragraph*{Introduction.} Fe$_2$P-based alloys have attracted significant research interest in recent years owing to their first-order magnetic transition (FOMT) coupled to a magnetoelastic transition, giving rise to a giant magnetocaloric effect in the vicinity of their Curie temperature \cite{PhysRevB.100.104439}, which is tunable across room temperature by suitable Fe-Mn and P-Si, P-B substitution \cite{GUILLOU2019403,doi:10.1002/adma.201304788}.
The latter, along with their composition by cheap and abundant elements, makes them eligible for energy transduction applications including solid-state  harvesting of thermal energy \cite{aenm.201903741, dzekan2020efficient} and real-case magnetocaloric refrigerators, that provide increased energy efficiency and substantial environmental benefits compared to gas compression thermodynamic cycles \cite{Bruck2005, Trung2010, DUNG2011, TEGUS2002}.

A FOMT is also shown by the parent compound Fe$_2$P \cite{WAPPLING1975258}, which however exhibits a much larger magneto-crystalline anisotropy (MCA) than Fe$_2$P-based Fe\-Mn\-P\-Si compounds \cite{Caron2013}, making it rather a candidate material  for permanent magnets. Indeed, its Curie temperature ($T_C \approx 220~\mathrm{K}$) is too low for most applications. However, $T_C$ can be raised well above room temperature by suitable Si, Ni, Co alloying while preserving a MCA nearly as large as in the parent compound \cite{PhysRevMaterials.1.051401}. It is therefore apparent that  pure \FP{}, though not directly applicable in magnetic or magnetocaloric technology, shares most of its physics with the derived alloys, while it is possibly a simpler system to model theoretically.

\FP\ crystallizes in the hexagonal C$_{22}$
structure with a space group $P\bar{6}2m ~(189)$ and the primitive unit cell contains three formula units and four inequivalent sites, with iron occupying the $3f$ (Fe1) and the $3g$ sites (Fe2) in equal number, and phosphorus occupying sites $2c$ (P1) and $1b$ (P2) in a 2:1 ratio \cite{CARLSSON197357, Scheerlinck1978}. The compound orders ferromagnetically (FM) with magnetic moments directed along the c-axis.
The magnetic structure of Fe$_2$P has been widely investigated by neutron scattering and M\"ossbauer spectroscopy \cite{WAPPLING1975258, Scheerlinck1978, Fujii1979, Tobola1996, KOUMINA1998, 10.1143/JPSJ.57.2143,PhysRevB.94.014426}.
All reports qualitatively agreed on a larger Fe2 moment with a localized character, and a reduced Fe1 moment typical of itinerant magnetism (a feature shared by FeMnPSi alloys \cite{DUNG2012}). However, poor quantitative agreement on the size of the Fe1 magnetic moment characterized early literature and, in addition, the presence of helical states below $T_{c}$ was discussed \cite{10.1143/JPSJ.57.2143}.
Recently, elastic neutron scattering experiments \cite{PhysRevB.99.174437} seem to have finally established the value of the Fe1 and Fe2 moments as 0.8~$\mu_{\mathrm B}$ and 2.11~$\mu_{\mathrm B}$, respectively. The same experiments also showed absence of canting below $T_{c}$ and the presence of sizable local moments on Fe up to 30~K above the FM transition temperature.

In this work we present an investigation of the magnetically ordered phase and of the magnetic transition of this compound by two local probes of magnetism, namely NMR and \msr{}.
Both techniques have been used to probe \FP\ only in their infancy and published results are very limited to the best of our knowledge \cite{WAPPLING1985347,10.1063/1.1660234}.

In zero applied field (ZF), $^{31}$P and $^{57}$Fe nuclei resonate in their hyperfine fields, giving rise to distinct resonance lines for each crystallographic site. We detected the $^{57}$Fe resonance of Fe2 and the $^{31}$P resonances of P1 and P2 and unambiguously assigned them to their respective nuclei, thus correcting the peak attribution by an early NMR work \cite{10.1063/1.1660234},
which is proven here to be erroneous. The so-determined $^{31}$P hyperfine fields effectively complement the determination of the  $^{57}$Fe hyperfine fields by M\"ossbauer spectroscopy \cite{WAPPLING1975258}.
ZF \msr{} showed a single sharp precession peak below $T_C$, whose low-temperature frequency poses stringent constraints to the  stopping site of the implanted muons,  while its temperature dependence confirms a FOMT in the system.
Experimental results are compared to a simulation of the system by  \emph{ab initio} methods, yielding theoretical predictions for the local fields at the $^{31}$P, $^{57}$Fe nuclei and at the muon in its stable interstitial site.

The motivation of this work is threefold. First,
the inconsistencies that can be found in the sparse and often very old literature on parent \FP{}, as pointed out above, demand clarification by newer experiments.
Second, this study will guide the interpretation of NMR and \msr{} experiments on \FP{}-based alloys of interest for applications.
Third, our results benchmark and validate \emph{ab initio} investigations that are shown to be extremely useful for experimental data analysis.

\paragraph*{Material and methods.}

\FP{} is prepared by firing a mixture of BASF carbonyl iron powder with red phosphorus under protective atmosphere. This mixture reacts exothermally and is very low in transition metal impurities, less than 0.01~\%. It was checked by X-ray diffraction to be single phase \FP{} type.

The NMR experiments were carried out by a home-built phase-coherent spectrometer \cite{Allodi2005} and a resonant LC probehead, using a field-sweeping cold-bore cryomagnet (Oxford Instruments Maglab EXA) equipped with a helium-flow variable temperature insert and a nitrogen-flow cryostat in zero field as sample environments at $T=5~\mathrm{K}$ and $T\ge80~\mathrm{K}$, respectively. ZF measurement at 77.3~K were performed by directly immersing the probehead in a liquid nitrogen dewar.  
Additional details about the experimental data acquisition are reported in the Supplemental Material (SM) \cite{}.

\msr{} experiments were performed on the LAMPF (Los Alamos Meson Physics Facility) spectrometer at TRIUMF in Vancouver, Canada and the General Purpose Surface-Muon Instrument at the Paul Scherrer Institut (PSI) in Villigen, Switzerland. The loose powder sample of \FP\ used for the experiment at TRIUMF was loaded into a mylar pouch and placed in a low-background sample holder. \msr{} spectra were collected in zero field (ZF) at temperatures between 100~K and 300~K using a helium flow cryostat to control the temperature. A calibration measurement was conducted in a weak transverse field at 275~K. The powder sample used at PSI was mixed with a small amount of wax. ZF spectra were collected at 5~K and 200~K using a helium flow cryostat. At both experimental facilities, data were collected in a warming sequence. Fits to the \msr{} spectra were conducted via least-squares optimization using MUSRFIT~\cite{SUTER201269}, and a home-built python package called BEAMS, both of which yielded statistically indistinguishable results for the oscillating frequencies.

The magnetic and structural stability of \FP~ and of various alloys have been already studied with different computational approaches \cite{SENATEUR1976631, DELCZEG2010,Bhat2018,PhysRevMaterials.1.051401}.
We reproduced previously reported results on the FM phase \cite{Ishida1987, Tobola1996, KOUMINA1998,Bhat2018} with both plane wave and full-potential approaches using the \textsc{Quantum ESPRESSO} suite of codes, the \WK{} package and the \elk{} code respectively \cite{Giannozzi2009, Giannozzi2017, elk, w2k, giannozzi2020}.
The plane wave and pseudopotential method is essential to reduce the computational effort required for muon site assignment and hyperfine field characterization, while full potential approaches provide superior accuracy for the evaluation of the hyperfine fields at P nuclei. All computational details are reported in the SM \cite{}.

\paragraph*{NMR}.
\begin{figure}
	\includegraphics[width=\columnwidth]{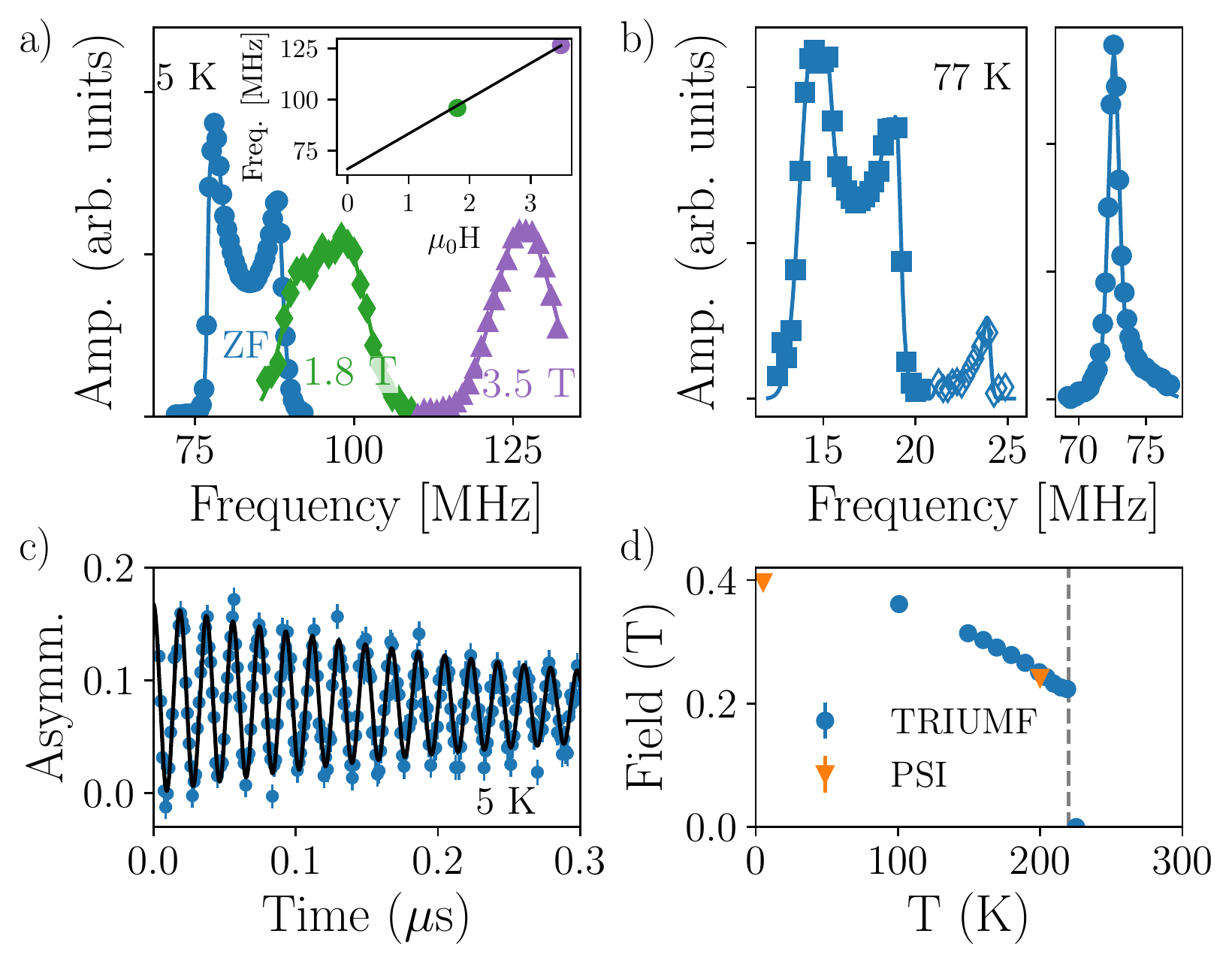}
	\caption{\label{fig:NMRmuSR} a) $^{31}$P(1) NMR spectra at $T=5~\mathrm{K}$ in ZF (domain wall signal) and in applied fields approaching saturation of the magnetization (domain signal).
    The inset shows the resonant frequency vs applied field, the black line is a fit described in main text.  b) ZF NMR spectrum at $T=77.3~\mathrm{K}$: $^{31}$P2 (left panel, filled squares), $^{57}$Fe2 (left panel, empty diamonds) resonance lines and $^{31}$P1 resonance (right panel, filled circles). In (a-b) the lines are guides to the eye.
    c) Zero-field \msr\ spectrum collected at 5~K. The blue dots represent the experimental data and the solid black curve is a fit obtained through least-squares optimization. The coherent oscillations indicate the presence of a well-defined, static magnetic field at the muon stopping site due to the long-range FM order. d) The static field at the muon stopping site as a function of temperature, obtained from fits to the ZF-\msr\ spectra. The vertical dashed line represents the Curie temperature ($\sim$220~K). }
	
\end{figure}
Spontaneous ZF NMR signals were detected at low temperature 
in the 70-90~MHz and the 13-24~MHz frequency intervals, in loose agreement with the frequency bands where similar resonances were reported by Koster and Turrell 
and assigned therein to $^{31}$P in  P1, P2 and $^{57}$Fe in Fe1 and Fe2, respectively \cite{10.1063/1.1660234}.

The higher-frequency portion of the ZF spectrum at 5~K is plotted in Fig.~\ref{fig:NMRmuSR}a. The relative high frequency indicates that these resonances stem from $^{31}$P nuclei (gyromagnetic ratio $^{31}\gamma/2\pi=17.2357~\mathrm{MH/T}$),
since a $^{57}$Fe resonance at the same frequency ($^{57}\gamma/2\pi=1.3786~\mathrm{MH/T}$) would correspond to  hyperfine fields of $\approx 60~ \mathrm{T}$, whence an  unphysical local moment $> 5\mu_B$ according to the known iron hyperfine coupling \cite{Novak_magnetite}. Nuclear spin echoes were excited with shorter and strongly attenuated rf pulses compared to standard NMR in non magnetic compounds, indicative of a large rf enhancement of the resonance \cite{Riedi_eta}. The mean enhancement factor $\eta$ was estimated in the order of 1000 by comparison with the optimum excitation conditions in an intense applied field saturating the magnetization (see below), where $1 < \eta < 2$ can be assessed \cite{Panissod}.
Such a large $\eta$ value prove that these signals originate from nuclei in domain walls \footnote{The enhancement factor of nuclei in the bulk of domains can be is estimated as $\eta_d = B_{hf}/B_{an} < 2$ \cite{Riedi_eta, Panissod} from the known value of the anisotropy field $B_{an}$, in the order of several tesla in  \FP \cite{Caron2013}.}.

The $^{31}$P ZF spectrum of Fig.~\ref{fig:NMRmuSR}a) exhibits a structure with two cusps at approx.\ 78 and 88~MHz. Such features were identified with independent resonance peaks and assigned to P2 and P1 by Ref. \cite{10.1063/1.1660234}.
However, such an attribution is inconsistent with experiments. 
First, the enhancement factors for the two spectral features differ by a factor of $\approx 3$ ($\eta$ is larger on the higher frequency side), a difference that cannot be reconciled with two close-frequency peaks in a homogeneous magnetic structure, and rather points to nuclei at different positions inside a domain wall \cite{Riedi_eta, Panissod}. Moreover, the two-cusp structure progressively disappears in an applied field large enough to saturate the magnetization. The same figure also shows the 5~K  spectra from $^{31}$P nuclei in the bulk of domains, in applied field values  $\mu_0 H'=1.8~\mathrm{T}$ and $\mu_0 H''=3.5~\mathrm{T}$ corresponding to a nearly saturated state and a practically full saturation of the magnetic moment of polycrystalline \FP{}, respectively \cite{Caron2013}. From $H'$ to $H''$, vector composition of the internal with the external field shifts the resonance frequency as $\delta\nu = \nu'' -\nu' =  \mu_0 (H''-H')^{31}\gamma/2\pi$, in agreement with $^{31}$P nuclei with a positive hyperfine coupling, while the lineshape tends to a single Gaussian curve at increasing field. It is therefore apparent that these NMR signals constitute a single resonance line, which is assigned to P1 by the DFT calculations detailed below.

The complex lineshape of the ZF spectrum at 5~K seemingly stems from  the anisotropic component of the hyperfine coupling and the particular (though unknown) micromagnetic structure of domain walls, whereby spins inside a wall do not sample the solid angle with equal probability. A uniform angle sampling relative to crystal axes, on the contrary, is approached by saturating the magnetization of the polycrystalline specimen. The linewidth of the spectrum in the larger applied field is estimated as $\Delta\nu=4.6~\mathrm{MHz}$, whence a rms anisotropic 
hyperfine field $B_{hf}^{(anis)} = \sqrt{3}\Delta\nu\, 2\pi/{^{31}\gamma} =0.46(1)~\mathrm{T}$. The isotropic hyperfine field at this site can be estimated from the first moment of the spectrum
$B_{hf}^{(iso)} =2\pi\bar \nu/{^{31}\gamma} =4.7(1)\mathrm{T}$, a result that however suffers from the uncertainty on the details of domain walls. A different estimate is obtained from the dependency of the resonant frequency on the applied field. The fit shown in the inset of Fig.~\ref{fig:NMRmuSR}a), where $\nu(0)$ is the only free parameter, provides $B_{hf}^{(iso)} = 2\pi \nu(0) /{^{31}\gamma} = 3.83(4)$~T.

At higher temperature, the $^{31}$P(1) ZF spectrum evolves to a single narrower peak. The two-cusp structure has already disappeared at 77.3~K (Fig.~\ref{fig:NMRmuSR}b), where only a weaker Gaussian shoulder can be detected besides the main peak at 72.6~MHz, and the overall spectral width is estimated as  $\Delta\nu=1.1(1)~\mathrm{MHz}$.
The narrowing of the ZF spectrum witnesses a decrease of the anisotropic hyperfine coupling of P1 from 5 to 77.3~K. On further warming, the mean resonance frequency vs temperature follows a smooth order parameter curve, with  $\bar\nu (T)$ values in good agreement with the literature, up to 160~K \cite{10.1063/1.1660234}. Above that temperature, the signal is lost due to exceedingly fast relaxations, and the magnetic transition cannot be probed by NMR.

Lower-frequency resonances, shown in Fig.~\ref{fig:NMRmuSR}b), were investigated at 77.3~K. The ZF NMR spectrum features a broad, more intense composite line at 14-18 MHz, and a weaker asymmetric peak at 23.7~MHz. The latter value is in excellent agreement with a $^{57}$Fe resonance in the hyperfine field of 17.2~T reported by $^{57}$Fe M\"ossbauer spectroscopy for Fe2 at this temperature \cite{WAPPLING1975258}, which warrants the same assignment for this NMR line.
The other broader resonance is however incompatible with the $^{57}$Fe NMR of Fe1, although such a resonance line is predicted  at 15.0~MHz ($B_{hf}=10.9~\mathrm{T}$) by the same M\"ossbauer data. In fact, the same integrated amplitude
(after normalization by $\nu^2$)
would be expected in that case for the two signals, given the 1:1 Fe occupancy ratio at the two sites. The 14-18 MHz signal must therefore originate from the resonance of the much more sensitive $^{31}$P nuclei in a mean spontaneous field of 0.94(4)~T at the complementary P2 site, while the  weaker and overlapped  $^{57}$Fe(1) line is hidden by it.
Our assignment, which contrasts with early literature \cite{10.1063/1.1660234}, is also further confirmed by the relative receptivity of the two nuclear species as detailed in the SM \cite{}.

The $^{31}$P(2) spectrum at 77.3~K exhibits a similar structure as the one observed at 5~K in the $^{31}$P(1) one, which can be explained based on similar arguments. Its linewidth and the P2  rms anisotropic hyperfine field  
are estimated as $\Delta\nu=1.8(1)~\mathrm{MHz}$ and $B_{hf}^{(anis)} = \sqrt{3}\Delta\nu\, 2\pi /{^{31}\gamma} =0.18(1)~\mathrm{T}$, hence they are significantly larger, both in absolute and relative terms, than the corresponding P1 values at the same temperature.

\paragraph*{\msr{}}. In Fig.~\ref{fig:NMRmuSR}c), we display the ZF-\msr\ spectrum collected at 5~K. 
The blue dots represent the experimental asymmetry as a function of time. Well-defined oscillations with a single dominant frequency are clearly visible, confirming the presence of a static and fairly uniform magnetic field at the muon stopping site. An excellent fit to the spectrum is obtained using the standard two-component model expected for static internal fields, consisting of an exponentially damped sinusoidal function and a slowly relaxing exponential function, as shown by the black curve in Fig.~\ref{fig:NMRmuSR}c). At 5~K, the best-fit frequency is $\nu=53.72(1)$~MHz, corresponding to a magnetic field magnitude of $B_{\mu} = \nu/\gamma_{\mu}=0.3963(1)$~T, where $\gamma_{\mu}=135.5~\mathrm{MHz~T^{-1}}$ is the gyromagnetic ratio of the muon. Equivalent fits were performed for the ZF spectra collected at temperatures up to $T_c\sim 220$~K, while above  $T_c$ a pure exponential decay is observed.

The local static field at the muon site extracted from these fits is displayed as a function of temperature in Fig.~\ref{fig:NMRmuSR}c), with blue circles and orange triangles representing the results from data collected at TRIUMF and PSI respectively. Excellent agreement is found between the TRIUMF and PSI results. As the temperature increases toward $T_c$, the static field steadily decreases, as expected for a magnetic order parameter curve.
At approximately 220~K (indicated by the vertical dashed line in Fig.~\ref{fig:NMRmuSR}d), however, the static field drops discontinuously to zero, indicating the occurrence of a first-order magnetic transition at this temperature. 
A fast depolarization (not shown) is observed well above $T_{c}$, indicating the presence of short range correlations in agreement with previous neutron scattering results.

\paragraph*{Computational results.}

In order to further characterize the microscopic origin of the experimental results and 
to validate \emph{ab initio} estimates of hyperfine couplings, we evaluated
the internal field at P and the muon sites, after
having identified the interstitial position occupied by the latter following a 
methodology already extensively discussed \cite{PhysRevB.87.115148, M_ller_2013, PbBonfa_2015, Onuorah_2018, PhysRevB.100.094401, JPSJ.89.014301,PERDEWPBE1996, MONKHORST1976,MARZARI1999,BLOCHL1994}.

\begin{table}[t]
\centering
\begin{tabular}{c | c  c  c  c }  
 \hline
 \hline 
    Label & Wyckoff & $(x,y,z)$ & $\Delta E(meV)$  & $B_C$ (T)\\ 
  \hline
   A & $3g $  &  (0.000,  0.328,  0.500)  &  0   & -0.4274\\
   A* & $6k $ &  (0.052,  0.358,  0.500)  &  0   & -0.5022 \\
   B & $3f $  &  (0.296,  0.296,  0.000)  &  280 & -0.4573  \\
   C & $2d $  &  (0.333,  0.666,  0.500)  &  690 & -1.7049 \\
   D & $3f$   &  (0.000,  0.545,  0.009)  & 760 & - \\
   E & $1a$   &  (0.000,  0.000,  0.000 ) & 1120 & - \\
 \hline
\end{tabular}
 \caption{Muon sites and contact hyperfine fields obtained \emph{ab initio}. The second and third columns report the position of the candidate muon sites in fractional coordinates with respect to a unit cell where Fe1 occupies the $3f$ position in (0.0,0.257,0.0) and Fe2 the $3g$ position in (0.0, 0.591, 0.5). The total energy differences $\Delta E=E_{i}-E_{A}$ and the contact hyperfine field at each site are in fourth and fifth columns.}
 \label{table:pos0}
\end{table}

\begin{figure}[tb]
\centering
\includegraphics[scale=0.5]{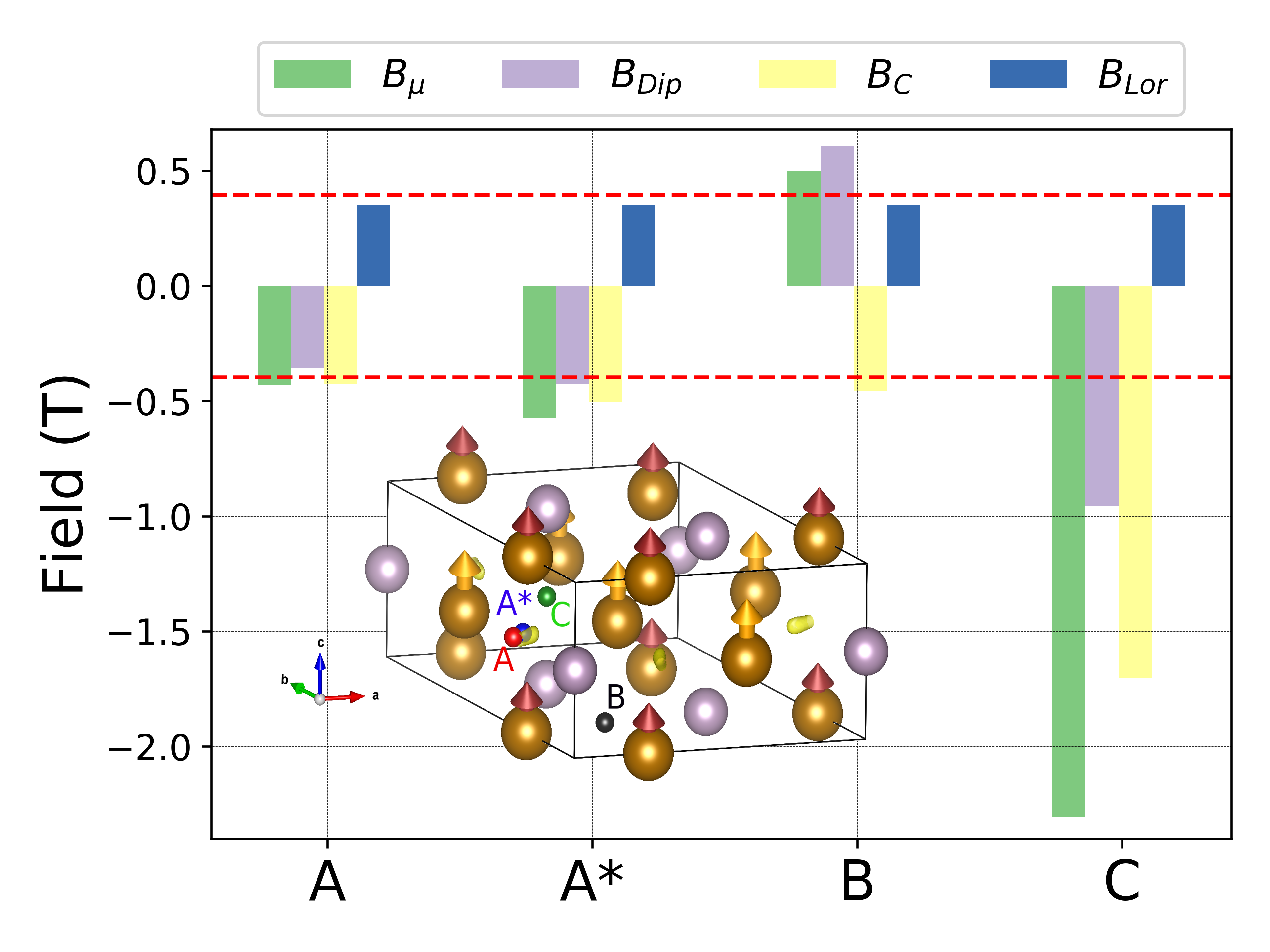}
\caption{Local field at the muon sites along the $\mathbf c$ lattice vector. The green bar is the sum of the LR dipolar (gray and blue), and Fermi contact (yellow) contributions. The red dashed line shows $\pm B_{\mu}$ (see text). The inset shows the \FP{} unit cell where on Fe atoms (brown) the short(long) red(orange) arrows identify the small(large) Fe1(Fe2) magnetic moments. P atoms are big mauve spheres and muon sites are the small spheres labeled A, A* , B and C. The yellow isosurfaces represent 0.2~eV above the minimum of the unperturbed electrostatic potential.}\label{fig:muonfields}
\end{figure}

Five inequivalent candidate muon sites, labeled with letters from A to E in order of increasing total energy, are reported in Tab.~\ref{table:pos0}. The label A* indicates a slightly displaced analogous of site A, with the distance $d_{A-A*}$ being 0.2~\AA.
The energy difference between the two is within numerical accuracy, but their distance testifies a rather flat potential energy surface that implies some degree of delocalization of the muon wave-function. 

Notably, the positions A and B are just $0.5$~\AA{} and 0.3~\AA{} away from the absolute minimum and the second lowest minimum of the electrostatic potential while site C corresponds exactly to the position of the third relative minimum of the electrostatic potential. A similar behavior was found in muon site calculations performed on FeCrAs that shares the same spacegroup with \FP \cite{Huddart_2019}. Finally, the largest displacement induced by the muon on the neighboring magnetic atoms is smaller than $0.15$\AA.

In a ZF NMR or \msr{} experiment performed below $T_C$, the effective field at the nuclei or muon can be separated into multiple contributions: 
\begin{align}
\mathbf{B}_{\mathrm{Tot}}&=\mathbf{B}_{\mathrm{LR}}+\mathbf{B}_{\mathrm{Demag}}+\mathbf{B}_{\mathrm{SR}} \label{eq:btot} \\
\mathbf{B}_{\mathrm{LR}} &= \mathbf{B}_{\mathrm{Lor}} + \mathbf{B}_{\mathrm{Dip}} \label{eq:blr} \\
\mathbf{B}_{\mathrm{SR}} &= \mathbf{B}_{\mathrm{C}} + \mathbf{B}_{\mathrm{DipSR}} + \mathbf{B}_{\mathrm{OrbSR}} \label{eq:bsr}
\end{align}

The first term in Eq.~\ref{eq:btot} is the long range (LR) dipolar field which is obtained here in real space
using the Lorentz method, in which the magnetic moment of Fe $3d$ orbitals inside a (large) sphere contribute to $\mathbf{B}_{\mathrm{Dip}}$ 
while those outside it are treated as a continuum and add into $\mathbf{B}_{\mathrm{Lor}}$ appearing in Eq.~\ref{eq:blr}.
$\mathbf{B}_{\mathrm{Demag}}$ is the demagnetization field, that can be neglected in a polycrystalline and multidomain sample, and $\mathbf{B}_{\mathrm{SR}}$ is the short range 
term arising from orbitals localized at the muon or P site, which is further subdivided, in order of appearance in Eq.~\ref{eq:bsr}, in the Fermi contact, dipolar and orbital terms.

For the LR dipolar field we approximate the spin polarization of Fe $d$ orbitals as classical magnetic moments with $m_{\mathrm{Fe1}}=0.84$ $\mu_{\mathrm{B}}$  and $m_{\mathrm{Fe2}}=2.22$ $\mu_{\mathrm{B}}$ both along $c$.

\begin{table}
\begin{threeparttable}

\begin{tabular}{| l |  c  | c c c || c | c |}
 \hline
   Code & Nuc. & $B_{\mathrm{C}}$ & $B_{\mathrm{DipSR}} $ & $  {B}_{\mathrm{Dip}} $ & $B^{\mathrm{(iso)}}_{\mathrm{exp}}$ & $\sqrt{5/2} B^{\mathrm{(anis)}}_{\mathrm{exp}}$\\
   \hline
\multirow{ 2}{*}{\WK} & P1  & 3.8  & -1.0 & + 0.11 & $3.83(4)^\dag$ & $0.73(2)\dag$ \\   
                         & P2  & 0.3  &  0.2 & - 0.19 & $0.94(4)^\star$ & $0.28(2)^\star$\\   
   \hline
   \multirow{ 2}{*}{\elk} & P1  & 4.0/3.2  & -0.8 & +0.11 &  $3.83(4)^\dag$ & $0.73(2)^\dag$ \\ 
                          & P2  & 0.2/0.7  &  0.1 & -0.19 & $0.94(4)^\star$ & $0.28(2)^\star$ \\ 
   \hline
 \end{tabular}
 \begin{tablenotes}
   \item  \small
   $^{\dag}$ data from applied field NMR; $^{\star}$ data from ZF NMR.
 \end{tablenotes}

 \caption{Hyperfine fields at the P nuclei in the low temperature FM phase. A positive contact field indicates that the spin polarization of a given contribution is oriented like Fe-$d$ orbitals, a negative sign means the opposite. The two values reported for $B_{C}$ in the second row are the results of two different algorithms used to compute the contact part (see SM). The fourth and fifth columns report the principal component of the dipolar tensor. All values are in Tesla and the Lorentz contribution to the long range dipolar field is 0.35~T. \label{tab:hyperfine}}
\end{threeparttable}
\end{table}

The occupied orbitals with non-negligible electronic density at the muon sites consist mainly of s-character and relativistic effects can be safely neglected. In this approximation, the short-range contribution (Eq.~\ref{eq:bsr}) is limited to the contact term, that is estimated from the electronic spin polarization at the muon position $\bm{R}_l$ as 
$ B_{\mathrm{SR}} = \frac{2}{3}\mu_{0}\mu_{B}\rho_{s}(\bm{R}_l)$ \cite{PhysRevB.47.4244,Onuorah_2018} where $\rho_{s}(\bm{R}_l)$ is the spin density at the muon site.
We can therefore compute $\mathbf{B}_{\mathrm{Tot}}(\bm{R}_l)$ entirely from first principles and the results are shown in Fig.~\ref{fig:muonfields} \cite{JPSCP.21.011052}.
The long range dipolar contributions is negligible for both the lowest energy sites A and B, where the local field originates from the Fermi contact term. The comparison with the experimentally measured local field is excellent and equally good for both sites, with the former showing slightly better agreement.

A similar approach is used to estimate the local field at P sites, where however relativistic effects must be considered \cite{PhysRevB.35.3271}. The contact term in this case dramatically depends on valence electrons' spin polarization and an accurate description of the latter is mandatory. 
The short-range dipolar and orbital contributions are estimated differently by \elk{} and \WK{} (see SM), but in both cases $B_{\mathrm{OrbSR}}$ is negligible and therefore not reported.

The calculated hyperfine field at P1 and P2 sites shown in Tab~\ref{tab:hyperfine} provides the attribution of the NMR peaks already discussed.
The contact term accurately reproduces the experimental bulk $B_{hf}^{(iso)}$ for P1 while the comparison for P2 is seemingly less accurate. However, an uncertainty of the order of the Lorentz term has to be considered since the experimental estimate for P1 is from domain wall signal and, notably, the difference between bulk and wall signals for P2 is 0.9~T.
The analysis of the anisotropic part requires more care.
In Tab~\ref{tab:hyperfine} the experimental $B_{hf}^{(aniso)}$ is multiplied by $\sqrt {5/2}$ in order to compare it with the largest principal value of the dipolar tensor.
In ZF, the uncertainty stemming from the unknown nature of the domain walls impairs a precise theoretical determination.
In applied field, at saturation, $B_{\mathrm{Lor}}$ and $B_{\mathrm{Demag}}$ cancel out and, for \FP, only $B_{\mathrm{DipSR}}$ and $B_{\mathrm{Dip}}$ contribute to
the anisotropic part. A good quantitative agreement is obtained in this case for P1.

\paragraph*{Conclusions.}
We investigated the FM phase of \FP{} using in-field and zero-field $^{31}$P NMR and \msr{}, characterizing in detail both experimental signals in
the low temperature magnetic phase.
First principles simulations unveiled the interstitial position occupied by the muon and provided hyperfine parameters for P nuclei and the muon, accurately reproducing the experimental results.
This information provides a framework for the analysis of future experiments on \FP{} alloys of technological interest.

\paragraph*{Acknowledgements.}
We thank Gerald Morris, Bassam Hitti, and Donald Arseneau for their support at the \msr\ beamline at TRIUMF, as well as Nicholas Ducharme, Alexander Shaw, Jacob Hughes, and Alec Petersen for their assistance collecting the data. We acknowledge the College of Physical and Mathematical Sciences at Brigham Young University for providing support to travel to TRIUMF. Analysis of the \msr\ data was supported by the United States Department of Energy, Office of Science, through award number DE-SC0021134.
We thank Francesco Cugini for fruitful discussion and we are in debt to John Kay Dewhurst and the developers of the \elk{} code for their prompt support and availability.
We acknowledge ISCRA C allocation at CINECA(Award no: HP10CJLG7W) and computer time allocations at SCARF and University of Parma.
This work is based on experiments performed at the Swiss Muon Source S$\mu$S, Paul Scherrer Institute, Villigen, Switzerland.
\bibliographystyle{apsrev4-2}
\bibliography{main}

\pagebreak
\widetext
\begin{center}
\textbf{\large Supplementary Material for ``\papertitle''}
\end{center}
\setcounter{equation}{0}
\setcounter{figure}{0}
\setcounter{table}{0}
\setcounter{page}{1}
\makeatletter
\renewcommand{\theequation}{SM\arabic{equation}}
\renewcommand{\thefigure}{SM\arabic{figure}}
\renewcommand{\thetable}{SM\arabic{table}}
\renewcommand{\bibnumfmt}[1]{[SM#1]}
\renewcommand{\citenumfont}[1]{SM#1}
\input{sm.tex}

\bibliographystyleSM{apsrev4-2}
\bibliographySM{SM}

\end{document}

%% file: sm.tex
\subsection{NMR}

The NMR spectra were reconstructed point by point by exciting spin echoes at discrete frequencies.
The corresponding spectral amplitude were determined as the zero-shift Fourier component of the echo signal,
divided by the frequency-dependent sensitivity $\propto\omega^2$.
At each frequency step, the LC resonator was re-tuned by a servo-assisted automatic system plugged into the spectrometer. A  standard $P-\tau-P$  spin echoes pulse sequence was employed, with equal rf pulses $P$ of intensity and duration optimized for maximum signal, and delay $\tau$ as short as possible, compatibly with the dead time of the apparatus.

A further check of the assignment of the 14-18 MHz signal to $^{31}$P nuclei, which contrasts with early literature \citeSM{SM_10.1063/1.1660234},
is obtained from the relative receptivity $R = R'B_{hf} = a\gamma^3B_{hf}^3 = a\omega^3 $ of the two nuclear species. Here $R' =  a\gamma^3B^2$ is the usual dependence of  the sensitivity of a nucleus on its abundance $a$ (both isotopic and from site multiplicity) and the local field $B$ in a non-magnetic substance, whereas in a ferromagnet a further dependency on $B=B_{hf}$ arises from the enhancement factor \citeSM{Riedi_eta, Panissod}.
After normalization of the spectra by $\omega^2$ (the amplitude correction appropriate for the signals of a single nuclear species), the integrated amplitudes $A$ of the $^{31}$P and $^{57}$Fe signals should scale relative to each other as $R_n =  a\omega$, whence an expected ratio ${}^{31}R_n / {}^{57}R_n = 11$, in fair agreement with the value ${}^{31}A / {}^{57}A \approx 16$ that we estimated experimentally. A direct comparison between the $^{31}$P NMR amplitudes at P1 and P2 is not possible due to the large difference in frequency whence the employment of different  resonant circuits.

The shape of the 77.3~K spectrum, testifying a variation of the anisotropic hyperfine coupling, was checked against different spin-echo excitation conditions (selecting nuclei with different enhancement factors)\citeSM{Panissod_Nato} and was found
 to be independent of rf pulse amplitude over more than two decades.

\subsection{DFT simulation details}
\label{sub:dft}

The magnetic ground state of \FP{} is accurately reproduced by the  Perdew-Burke-Ernzerhof (PBE) exchange and correlation functional \citeSM{SM_PERDEWPBE1996}. The energy difference between the 
ferromagnetic and the spin-degenerate case is $1.1$~eV  per unit cell and the magnetic moment of Fe1 is found to be 0.84~$\mu_{\mathrm B}$ while that of Fe2 is 2.22~$\mu_{\mathrm B}$. Both these results are in agreement with previous computational studies and compare extremely well with experimental findings.\citeSM{SM_Tobola1996, SM_KOUMINA1998, SM_Ishida1987, SM_Bhat2018}

The lattice structure of \FP ~ was taken from the crystallographic information file deposited on the COD database \citeSM{SM_Quiros2018} with ID: 9012616.

In the simulations using plane wave and pseudopotential, the basis set was expanded up to a cutoff of 80~Ry while charge density up to 800~Ry.
A uniform Monkhorst-Pack (MP) \citeSM{SM_MONKHORST1976} mesh of $6\times\ 6 \times 8$ was used for sampling the reciprocal space in the unit cell while we used a $2a\times2b\times3c$ supercell to identify muon interstitial sites.

In the optimization of the structure and description of magnetic properties a Marzari-Vanderbilt \citeSM{SM_MARZARI1999} smearing of 5 mRy was used while a Gaussian smearing of 20 mRy is chosen for supercell calculations.

The experimental lattice parameters $a = b = 5.877$ and $c = 3.437$\AA{} \citeSM{SM_SENATEUR1976631, SM_CARLSSON197357, SM_Scheerlinck1978} have been used throughout all the calculations.
The optimized atomic positions show very small displacements , always smaller than $5\cdot 10^{-2}$~\AA. In order to further validate our description we also optimized lattice parameters and found $a = b = 5.815$ and $c = 3.425$~\AA.

In all cases we performed spin-polarized calculations using the generalized gradient approximation (GGA) proposed by Perdew Burke and Ernzerhof (PBE)\citeSM{SM_PERDEWPBE1996} and Projector Augmented Wave (PAW) \citeSM{SM_BLOCHL1994} pseudopotentials that are required to reconstruct the core electrons polarization.

For muon site assignment, the set of initial muon locations and the \FP~ atomic positions were fully relaxed in a $2a\times 2b\times 3c$ supercell until forces and energy difference between optimization steps were smaller than $10^{-3}$~Ry/a.u and $10^{-4}$~Ry. The full list of positions used to perform the calculation is in Tab.~\ref{table:site_positions}.

A $4\times4\times4$  uniform grid is used to sample the interstitial voids in the unit cell. The number points actually used as starting guess for the muon site reduces to 8 when positions closer than 1~\AA ~ to hosting atoms and symmetry equivalent points have been removed. Three additional positions were identified from the disconnected minima of the bulk electrostatic potential and used as an additional starting guess.
\begin{table}[h!]
\centering
 \caption{Summary of candidate muon stopping sites. The first 8 sites are from $4\times4\times4$ uniform grid and the last 3 from the bulk electrostatic potential.}
  \label{table:site_positions}
   \begin{threeparttable}
\begin{tabular}{ c | c | c | c | c  }  
\noalign{\smallskip} \hline \hline \noalign{\smallskip}
   & Wyckoff\tnote{a} & $(x,y,z)$\tnote{b} & $(x,y,z)$\tnote{c} & $\Delta E(eV)$\tnote{d} \\ 
  \hline
   1 & $1a $  & 0.00  0.00  0.00 & 0.0000  0.0001  0.0001  & 1.1671 \\
   2 & $3g $  & 0.00  0.25  0.50 & 0.0024  0.3274  0.5000  & 0.0006 \\
   3 & $3f $  & 0.00  0.50  0.00 & 0.0000  0.5473  0.0074  & 0.7464 \\
   4 & $6i $  & 0.00  0.50  0.25 & 0.0002  0.5314  0.1038  & 0.7344 \\
   5 & $3f $  & 0.00  0.75  0.00 & 0.0000  0.7068  0.0003  & 0.2545 \\
   6 & $6i $  & 0.00  0.75  0.25 & 0.0000  0.7054  0.0160  & 0.2556 \\
   7 & $12l$  & 0.25  0.50  0.25 & 0.0578  0.3595  0.4997  & 0.0009 \\
   8 & $6k $  & 0.25  0.50  0.50 & 0.0486  0.3535  0.5000  & 0.0000 \\
   9 & $6k $  & 0.10  0.35  0.50 & 0.0554  0.3572  0.4999  & 0.0003 \\
  10 & $3f $  & 0.35  0.35  0.00 & 0.2959  0.2959  0.0038  & 0.2549 \\
  11 & $2d $  & 0.33  0.66  0.50 & 0.3320  0.6654  0.5001  & 0.6324 \\
\noalign{\smallskip} \hline \noalign{\smallskip}
\end{tabular}
    \begin{tablenotes}
    \item [a] wyckoff number \label{taa}
    \item [b] starting position
    \item [c] final relaxed position
    \item [d] total DFT energy difference w.r.t lower energy site
    \end{tablenotes}
 \end{threeparttable}
 \end{table}
Finally, for the estimation of the contact hyperfine field at the muon sites, the reciprocal space sampling was increased to a $4\times 4\times 4$ Monkhorst-pack mesh grid.

For all electron simulations performed with Elk, in order to achieve convergence of hyperfine fields estimates, a dense $10\times10\times12$ MP reciprocal space grid is used while
atomic positions were kept fixed in the experimentally reported values. For simulations with spin-orbit contribution, the k-point grid was reduced to $7\times7\times9$, to speedup the simulation at a negligible loss of accuracy on the scale of the uncertainties discussed below.
A plane-wave cutoff of $\left | G + K \right |_\text{max} = 9/R_\text{min}^{MT}$ ($R_\text{min}^{MT}$ is the average of the muffin-tin radii in the unit cell) was used for the expansion of the wavefunction in the interstitial region.
The muffin-tin radius for Fe and P are $1.98$ and $2.17$ a.u. and respectively.
The number of empty states was increased until it reached 40\% of valence states and spin-orbit coupling was considered.
For Fe, we moved semi-core $s$ states to the core and treated them with the full Dirac equation including core polarization.
In the Elk code, the Dirac equation is solved with the collinear spin formalism and the magnetization is assumed to be directed as the valence magnetization.

The results from WIEN2k code were obtained with version 18.2. In this case the reciprocal space grid was $12\times12\times18$ and muffin-tin radii were $2.35$ and $1.83$ a.u. for Fe and P respectively. The plane waves were expanded up $R_{MT}k_{max}=8$ where $R_{MT}$ is the average radius of MT spheres while non-spherical potential and charge density inside MT were expanded up to $l=10$. A total charge difference smaller than $10^{-6}e$ has been set as SCF convergence criterion.

\begin{table}
\begin{tabular}{| c |  c  | c | c | c |}
 \hline
   Code & Nucleus & Cont. & Dipolar SR & Orbital SR \\
   \hline
   \multirow{ 4}{*}{ Elk} & P1   &  4.0(5)   & -0.8 & 0.1 \\
                          & P2   &  0.2(5)   &  0.1 & 0.1 \\
                          & Fe1  & -15.8(5)  &  0.6 & 1.5 \\
                          & Fe2  & -17.2(6)  & -0.9 & 4.6 \\

   \hline
   \multirow{ 4}{*}{ Wien2K } & P1   &  3.8   & -1.0 & 0.0 \\
                              & P2   &  0.3   &  0.2 & 0.0 \\
                              & Fe1  & -13.1  &  0.5 & 1.5 \\
                              & Fe2  & -14.9  & -0.7 & 4.3 \\

   \hline
 \end{tabular}
 \caption{Hyperfine field at the P and Fe nuclei obtained with Elk and Wien2K with magnetic moments parallel to the $\mathbf c$ lattice vector. All values are in Tesla and are obtained from simulations including spin orbit interaction. The numbers appearing in the brackets are the standard deviation of the two estimates of the contact part obtained with the two different methods implemented in Elk.}\label{tab:hyperfinesign}
\end{table}

\subsection{Hyperfine Fields}

Following Philippopoulos \emph{et al.} in Ref.~\citeSM{SM_PhysRevB.101.115302}, we write the hyperfine interaction between the electrons and the nuclear or muon spin $\bm{I}_l$ at position $l$ as 
\begin{equation}\label{eq:Hhf}
\mathcal{H}_{\mathrm{hf}}= -\hbar \gamma_{l} \bm{h}_l\cdot\bm{I}_l
\end{equation}
where, $\gamma_{l}$ is the gyromagnetic ratio of the muon or the nuclear isotope $l$ and $\bm{h}_l$ is the hyperfine field operator.  

In magnetic materials, both core and valence electrons contribute to the hyperfine field, with the former requiring the solution of the Dirac equation to correctly account for relativistic effects in heavy nuclei. 
The hyperfine field in localized magnetic system can be split into a contribution from the electronic density surrounding the spin of interest within a typical atomic dimension (short-range contribution) and a contribution from electron density localized at distant sites (long-range contribution). The latter can be described in the classical limit by treating the distant spin polarized electronic orbitals as classical magnetic dipoles.

Let us focus on the short-range interaction first. By tracing out the electronic degrees of freedom, one obtains the matrix elements of $\bm{h}_l$ that describes the short-range contributions to the hyperfine field for atom or muon $l$ at position $\bm{R}_l$. These are
\begin{eqnarray}
\mathbf{h}^{l}_{\nu\nu'}  & = & \int_\Omega d^3 r \psi^\dagger_\nu(\mathbf{r})\mathbf{h}(\mathbf{r}-\bm{R}_l)\psi_{\nu'}(\mathbf{r}),\label{eq:hMatrixElements}\\
\mathbf{h}(\mathbf{r}) & = & \frac{\mu_0}{4\pi}\left(2\mu_B\right)\left(\frac{\boldsymbol{\sigma}}{2}\cdot\overleftrightarrow{T}(\mathbf{r})+\sigma_0\frac{1}{r^3}f_\mathrm{T}(r)\mathbf{L}\right),\label{eq:hrSingleParticleMatrix}\\
f_\mathrm{T}(r) & = & \frac{r}{r+r_\mathrm{T}/2}\label{eq:fT},
\end{eqnarray}
where the electronic wave-functions are approximated using Kohn-Sham single particle wavefunctions $\psi_{\nu}(\mathbf{r})$, $\mu_0$ is the vacuum permeability, $\mu_\mathrm{B}$ is the Bohr magneton, the electron $g$-factor is $g\simeq 2$, $\boldsymbol{\sigma}$ are Pauli matrices with $\sigma_0$ being the identity.
The second term in Eq.~\eqref{eq:hrSingleParticleMatrix} is the contribution originating from electron angular momentum, and
the factor $f_\mathrm{T}(r)$ accounts for a cutoff at short distances on the order of the Thomson radius $r_{\mathrm T}$.
The first term in Eq.~\ref{eq:fT} includes both the Fermi Contact and the spin dipolar interaction, and has the explicit form:
\begin{eqnarray}\label{eq:tensor}
T^{\alpha\beta}(\mathbf{r})&=&\frac{8\pi}{3}\delta_\mathrm{T}(\mathbf{r})\delta_{\alpha\beta}+\frac{3 r_\alpha r_\beta-r^2\delta_{\alpha\beta}}{r^5}f_\mathrm{T}(r),\label{eq:Ttensor}\\
\delta_\mathrm{T}(\mathbf{r})&=&\frac{1}{4\pi r^2}\frac{df_\mathrm{T}(r)}{dr},\label{eq:deltaT}
\end{eqnarray}
where $\alpha,\beta \in\{x,y,z\}$. 

The short-range contributions are estimated using different approaches in WIEN2k and Elk. WIEN2k exploits the fact that inside the muffin-tin the basis functions are atomic-like and the expectation values of the operators in Eq.~\ref{eq:hMatrixElements} at the various nuclear positions can be easily computed if the interstitial part is neglected. The contact part is finally estimated with an average over the Thomson radius of the spin density at the nucleus.
The Elk code computes instead the total field at nuclear sites by solving the Poisson equation for the vector potential where the current density is obtained from the electronic core and valence spin magnetization density and, optionally, from the orbital current density. The value is eventually averaged over the nuclear radius
\citeSM{SM_ANGELI2004185}. When the orbital contribution is neglected, the isotropic part of the hyperfine tensor obtained in this latter method represents an alternative estimate of the contact hyperfine field and the standard deviation obtained from the two values is used to quantify the accuracy of the computational estimate in Tab.~\ref{tab:hyperfinesign}.

Finally, the long range interaction is obtained with the approach proposed by Lorentz which involves the sum of two contributions: a real space sum of the field produced by the magnetic dipoles inside a large sphere of radius $r_{\mathrm L}$ and a second contribution due to magnetic moments outside the sphere. These simulations have been converged against the size of $r_{\mathrm L}$ and a $100\times100\times200$ supercell was constructed to host the sphere.